\documentclass[aip,apl,superscriptaddress,reprint]{revtex4-1}

\usepackage{graphicx,amsmath,amssymb,color}
\usepackage{tabularx, ctable}

\usepackage{bm}
\usepackage{siunitx}
\usepackage{gensymb}
\usepackage{flushend}

\usepackage[breaklinks=true,colorlinks,allcolors=blue]{hyperref}

\newcommand{\be}{\begin{eqnarray}}
\newcommand{\ee}{\end{eqnarray}}

\begin{document}

\title{Room-temperature graphene sub-terahertz detector integrated on a silicon dielectric waveguide}

\author{A. Titchenko}
\email{anlyubchak@miem.hse.ru}
\affiliation{National Research University Higher School of Economics, Moscow, 101000}
\affiliation{Moscow Pedagogical State University, Moscow 119991}

\author{K. Shein}
\affiliation{National Research University Higher School of Economics, Moscow, 101000}

\author{M.~Titova}
\affiliation{Programmable Functional Materials Lab,
Center for Neurophysics and Neuromorphic Technologies, Moscow, 121205, Russia}
\affiliation{Center for Advanced Studies, Kulakova str, Moscow}

\author{M.~Kashchenko}
\affiliation{Programmable Functional Materials Lab,
Center for Neurophysics and Neuromorphic Technologies, Moscow, 121205, Russia}
\affiliation{Center for Advanced Studies, Kulakova str, Moscow}
\author{O. Popova}
\affiliation{Programmable Functional Materials Lab,
Center for Neurophysics and Neuromorphic Technologies, Moscow, 121205, Russia}
\affiliation{Center for Advanced Studies, Kulakova str, Moscow}
\author{R.~Izmaylov}
\affiliation{National Research University Higher School of Economics, Moscow, 101000}
\author{E.I. Titova}
\affiliation{Programmable Functional Materials Lab,
Center for Neurophysics and Neuromorphic Technologies, Moscow, 121205, Russia}
\affiliation{Center for Advanced Studies, Kulakova str, Moscow}

\author{I. Gayduchenko}
\email{igaiduchenko@hse.ru}
\affiliation{National Research University Higher School of Economics, Moscow, 101000}

\author{G. Goltsman}
\affiliation{Russian Quantum Center, 30 Bolshoy Boulevard, building 1, Moscow, 121205, Russia}
\affiliation{National Research University Higher School of Economics, Moscow, 101000}
\affiliation{Moscow Pedagogical State University, Moscow 119991}
\begin{abstract}

Terahertz (THz) photonics provides a compelling foundation for next-generation 6G wireless communications, on-chip spectrometers, and non-invasive biomedical sensors. In the near- and mid-infrared, monolithically integrated photodetectors are a cornerstone of photonic integrated circuits (PICs). In the THz range, however, such integration on low-loss dielectric waveguides remains an outstanding challenge: to date, room-temperature on-waveguide detection has been realized only through hybrid flip-chip assembly of discrete elements, which degrades mode coupling, increases device footprint, and compromises mechanical stability. Here we address this gap by demonstrating a room-temperature graphene sub-THz detector monolithically integrated on a high-resistivity silicon dielectric waveguide for the D-band (110–170 GHz) operation. We couple an hBN-encapsulated graphene channel to the guided mode using a tapered slot-line antenna patterned directly onto the silicon surface. The detector achieves a voltage responsivity of $\sim$11.5 V/W and a 3 dB bandwidth of 1.94 GHz, with the latter currently limited by parasitic inductance of the readout interconnects rather than by the intrinsic graphene response. This platform provides a scalable route toward fully integrated room-temperature THz photonic circuits, where electrostatic gating and impedance matching offer clear pathways to order-of-magnitude improvements in both responsivity and bandwidth.

\end{abstract}

\maketitle
\begin{figure*}[!t]
  \centering
  \includegraphics[width=0.95\linewidth]{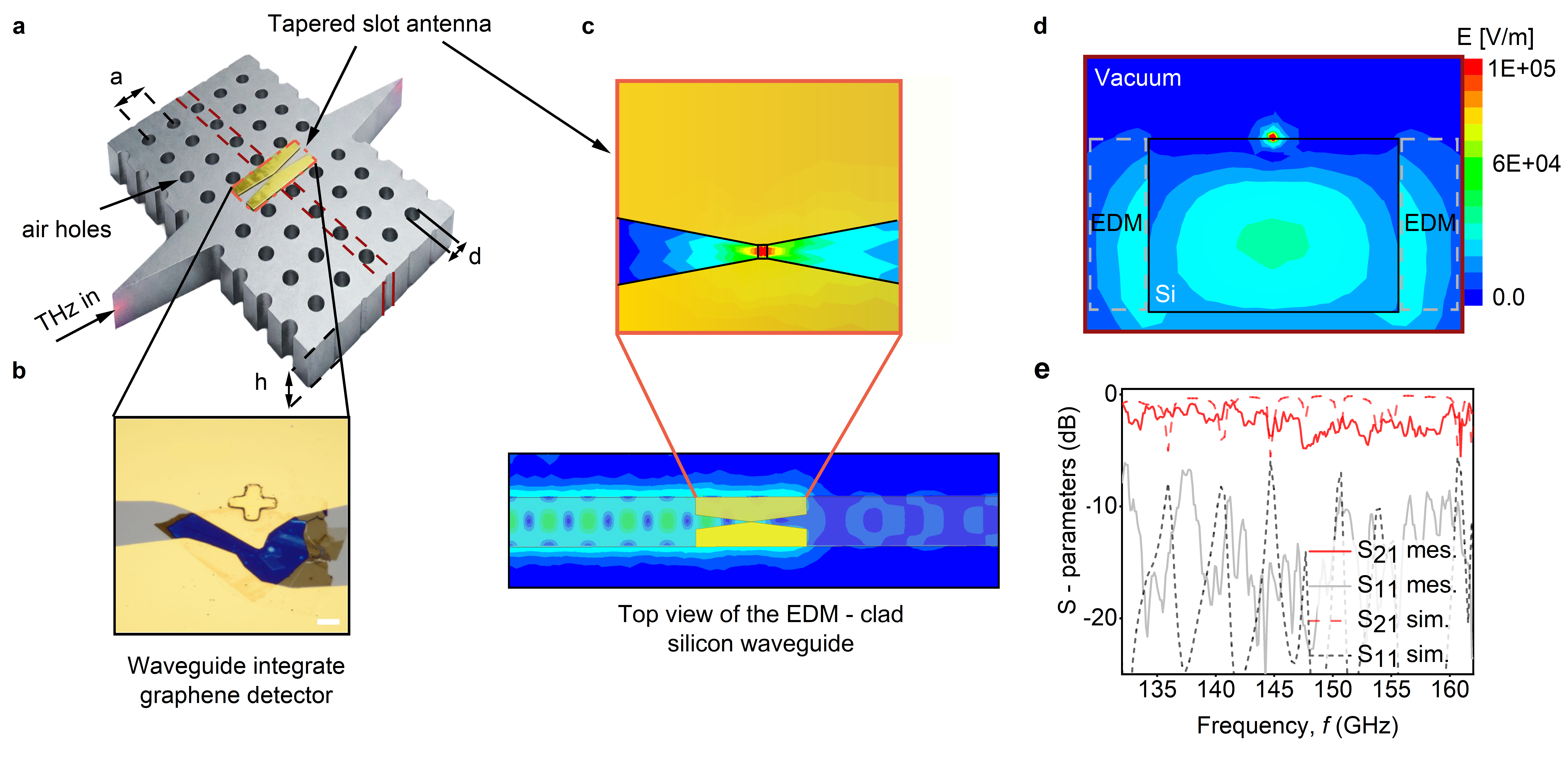}
  \caption{
    \textbf{Design and electromagnetic characterization of the graphene-on-silicon-waveguide sub-terahertz detector.}
    \textbf{(a)} High-resistivity silicon dielectric waveguide ($\rho > 3~\mathrm{k\Omega\cdot cm}$, thickness $h = 400~\mu\mathrm{m}$) with effective-dielectric-medium (EDM) cladding formed by a square lattice of subwavelength through-holes (period $a = 165~\mu\mathrm{m}$, hole diameter $d = 73~\mu\mathrm{m}$). An hBN-encapsulated graphene detector is monolithically integrated at the waveguide center and coupled to the guided mode via a tapered slot-line antenna. 
    \textbf{(b)} Optical micrograph of the antenna-coupled graphene detector (active channel $10 \times 20~\mu\mathrm{m}$, scale bar: 10 $\mu$m)
   \textbf{(c)} Simulated $|E|$ at 139~GHz: in-plane distribution showing concentration at the feed gap (top) and longitudinal propagation showing attenuation past the detector (bottom).
    \textbf{(d)} Cross-sectional mode profile, where the EDM cladding is represented by a homogeneous dielectric with effective permittivities $\varepsilon_{\mathrm{TE}}$ and $\varepsilon_{\mathrm{TM}}$ derived from the hole-lattice geometry. The fundamental guided mode is confined within the silicon core with an evanescent tail extending to the graphene layer on the waveguide surface.
    \textbf{(e)} Measured (solid) and simulated (dashed) $S$-parameters of the bare waveguide (132--162~GHz).
}
  \label{Fig1}
\end{figure*}

The field of THz and sub- THz technologies has advanced rapidly over the past decade, driven by applications in 6G wireless communications, radio astronomy, security imaging, high-resolution spectroscopy, and non-invasive biomedical diagnostics~\cite{sarieddeen2021overview,jiang2021road,amini2021review,li2025terahertz}. Sustained progress in these areas requires compact and energy-efficient solutions for the generation, detection, and on-chip manipulation of radiation in the sub-THz and THz range. The mature silicon and III–V near-infrared photonic platforms~\cite{Schall2014} provide the foundation for terahertz photonic integrated circuits (PICs)~\cite{Nagatsuma2016,Sengupta2018,nagatsuma2013terahertz,kang2024frequency}. Monolithically combining sources, detectors, and passive routing elements on a single substrate via low-loss dielectric waveguides~\cite{siew2021review,jiang2025advances,leuthold2010nonlinear}, these PICs represent a leading architecture for next-generation THz systems.

The passive component toolkit of THz PICs---waveguides, couplers, phase shifters, resonators, beam-forming networks---is now relatively mature~\cite{headland2025robust,jia2023valley,Seliverstov2023,Seliverstov2025,deng2022chip,gupta2024150,gao2025ultra,dechwechprasit20231,zhou2021photonics,rui2024all,lees2024terahertz,wang2024chip}. In contrast, the monolithic integration of active elements --- detectors, sources~\cite{headland2022terahertz}, and modulators~\cite{zhang2026chip,herter2023terahertz} --- on the same dielectric waveguide platform remains an outstanding challenge. Reports of detectors integrated directly on a THz waveguide are scarce~\cite{Shurakov2023,Yu2020,TorresGarcia2020,ichikawa2025terahertz,suminokura2014integration,yu2018integrated,nagatsuma2025inp,torres2020silicon,okamoto2017terahertz} and fall into two categories with complementary limitations. The first relies on superconducting bolometric elements monolithically defined on the waveguide~\cite{Shurakov2023}. Such devices offer potentially high sensitivity but require operation at $T \approx 4.2$~K, which precludes deployment in compact, room-temperature systems. The second relies on post-fabrication hybrid assembly of discrete detector or mixer chips onto the THz waveguide~\cite{ichikawa2025terahertz,suminokura2014integration,yu2018integrated,nagatsuma2025inp,torres2020silicon,okamoto2017terahertz}. This approach introduces alignment-sensitive mode mismatch and reflections at the chip-to-component interface, and it adds serial pick-and-place steps that are poorly suited to a monolithic, wafer-scale process flow.

A monolithically integrated, room-temperature photodetector compatible with low-loss dielectric THz waveguides is therefore still missing. Graphene is a natural candidate to fill this gap. It can be transferred onto arbitrary substrates by mature dry-transfer techniques, its Fermi level is electrostatically tunable, and its weak electron-phonon coupling underlie an efficient bolometric response at $T = 300$~K~\cite{yan2012dual,yuan2020room,sassi2017graphene}. Free-space graphene THz and sub-THz detectors have already been demonstrated~\cite{Castilla2019,zhou2026gate,gayduchenko2021tunnel,bandurin2018resonant,bandurin2018dual,gayduchenko2018manifestation,viti2025scalable,soundarapandian2026high,thomson2024coherent,ludwig2024terahertz,caridad2024room}, motivating their integration into a fully on-chip architecture.

In this work, we demonstrate a room-temperature, monolithically integrated sub-THz detector based on an hBN-encapsulated graphene channel on a high-resistivity silicon dielectric waveguide, operating in the D-band ($110$--$170$~GHz) at $f_{0} = 139$~GHz. The graphene channel is coupled to the guided mode through a tapered slot-line antenna patterned directly on the silicon surface. The device achieves a voltage responsivity $R_{V} = 11.5$~V/W and a $3$-dB bandwidth $f_{3\mathrm{dB}} = 1.94$~GHz. To the best of our knowledge, this is the first room-temperature detector monolithically integrated onto a silicon dielectric THz waveguide~\cite{Shurakov2023}.

The device architecture and electromagnetic characterization are summarized in Fig.~\ref{Fig1}. The platform is based on a high-resistivity silicon dielectric waveguide (Fig.~\ref{Fig1}(a)) designed for low-loss D-band ($110$--$170$~GHz) operation. The waveguide employs an effective-dielectric-medium (EDM) cladding~\cite{Tsuruda2015,gao2019effective,headland2023terahertz} formed by a square lattice of subwavelength through-holes (lattice period $a = 165$~$\mu$m, hole diameter $d = 73$~$\mu$m), which provides transverse confinement via total internal reflection while maintaining broadband, low-dispersion characteristics. In the long-wavelength limit, this perforated cladding is described as a homogeneous anisotropic medium with polarization-dependent effective permittivities
\begin{equation}
   \varepsilon_{\mathrm{TE}} = \varepsilon\,\frac{1 + \zeta + \varepsilon(1 - \zeta)}{1 - \zeta + \varepsilon(1 + \zeta)}, \qquad \varepsilon_{\mathrm{TM}} = \zeta + \varepsilon(1 - \zeta)\,,
    \label{eq:eps_eff}
\end{equation}
\begin{figure*}[ht!]
  \centering\includegraphics[width=0.9\linewidth]{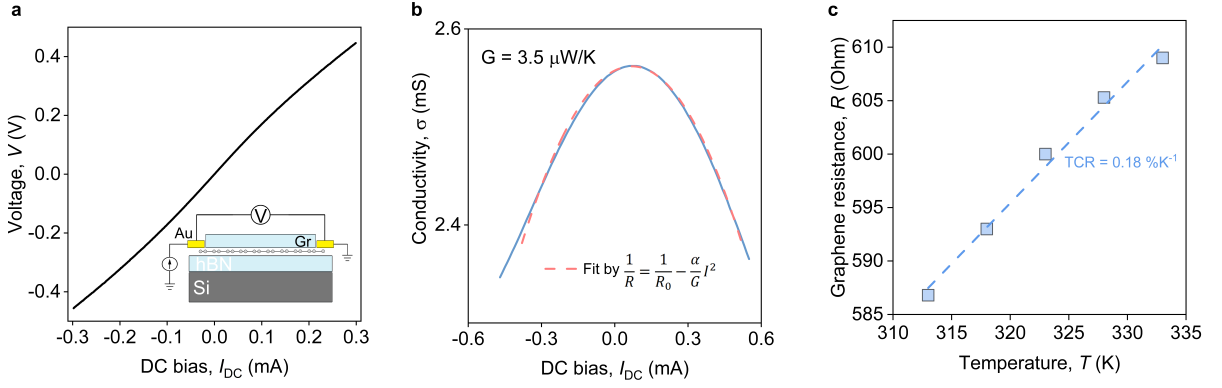}
    \caption{\textbf{Electrical transport characterization of the graphene channel at room temperature.}
   \textbf{(a)} Current--voltage characteristic of the hBN-encapsulated graphene channel measured at $T_{0} = 300$~K. The trace is symmetric in bias polarity and weakly nonlinear, the deviation from ohmic behavior at higher $|I_{\mathrm{dc}}|$ arising from Joule self-heating. Inset: schematic cross-section showing Au contacts to the graphene channel encapsulated between top and bottom hBN flakes on the silicon waveguide.
    \textbf{(b)} Differential conductivity $\sigma = dI/dV$ as a function of DC bias current $I_{\mathrm{dc}}$ (solid blue), together with the fit to the self-heating model $1/R(I) = 1/R_0 - (\alpha/G) I^{2}$ (dashed red). The fit yields a thermal conductance $G = 3.5$ $\mu$W/K (with $\alpha$ fixed from panel (c)).
    \textbf{(c)} Temperature dependence of the differential resistance $R_d$ measured between 313 and 333 K (symbols), with a linear fit (dashed line) yielding a temperature coefficient of resistance TCR $= (1/R_0) (dR/dT) = 0.18\%$ K$^{-1}$.
    }
	\label{Fig2}
\end{figure*}
where $\varepsilon = 11.9$ is the silicon permittivity and $\zeta = \pi d^{2}/(4 a^{2}) = 0.158$ is the air filling fraction of the hole lattice. For the chosen geometry this yields $\varepsilon_{\mathrm{TE}} = 9.10$ and $\varepsilon_{\mathrm{TM}} = 10.17$. Input/output coupling to WR-6 rectangular waveguides is achieved through adiabatic tapered transitions with half-angle $\alpha = 10^{\circ}$. The silicon core has a thickness $h = 400$~$\mu$m and a width of $580$~$\mu$m, dimensions chosen so that the waveguide supports a single guided mode across the operating band. Measured and simulated $S$-parameters of the bare waveguide, shown in Fig.~\ref{Fig1}(e) over $132$--$162$~GHz, confirm low reflection and transmission losses.

The graphene detector is monolithically integrated at the center of the waveguide (Fig.~\ref{Fig1}(b)). It comprises an hBN‑encapsulated graphene channel ($  L \times W$ = 10 $\times$ 20 $\mu$m) placed across the feed gap of a tapered slot‑line antenna patterned on the silicon surface. As in the near-infrared and visible ranges~\cite{Pospischil2013,Schall2014,Wang2014}, the graphene layer absorbs THz radiation by direct coupling to the evanescent waveguide mode. At 139~GHz the in-silicon wavelength ($\lambda_{Si} = 0.63$~mm) would normally require a large absorbing area, but this raises the heat capacity and degrades bolometric sensitivity. To resolve this trade-off, the tapered slot-line antenna~\cite{Shurakov2023,lyubchak2024waveguide} focuses the guided wave onto the compact channel, enabling efficient absorption while keeping the heat capacity low.

Full-wave finite-element simulations confirm efficient coupling of the guided mode to the graphene channel. In the model, the perforated EDM cladding is represented by a homogeneous anisotropic medium with the effective permittivities $\varepsilon_{\mathrm{TE}}$ and $\varepsilon_{\mathrm{TM}}$ of Eq.~\eqref{eq:eps_eff}. This representation is valid in the long-wavelength regime, where the lattice period $a$ is well below the in-silicon wavelength $\lambda_{Si}$ at 139~GHz and the hole lattice acts as an effective continuous medium. At $f = 139$~GHz, the in-plane electric-field distribution exhibits a pronounced maximum localized between the two metallic lobes of the slot-line antenna, that is, at the feed gap where the graphene channel bridges the two contacts (Fig.~\ref{Fig1}(c), top). The appearance of this field maximum precisely at the antenna feed, rather than along the metal edges or in the bulk silicon, is a direct signature that the antenna concentrates the guided-mode energy onto the active graphene area.  Simulated field propagation along the full device length at $139$~GHz (Fig.~\ref{Fig1}(c), bottom) shows clear attenuation of the mode past the detector position, confirming energy extraction from the guided wave into the graphene channel by the antenna. The cross-sectional view (Fig.~\ref{Fig1}(d)) reveals that the fundamental guided mode is mainly confined within the silicon core but possesses a pronounced evanescent tail extending above the waveguide surface. This evanescent field overlaps with the graphene layer, enabling efficient absorption of the guided sub-THz radiation. This coupling is quantified by the simulated three-port response of the antenna-coupled detector. In the three-port model, ports~1 and~2 correspond to the two ends of the silicon waveguide and port~3 to the coaxial feed of the slot-line antenna where the graphene channel is located. At $139$~GHz the reflection at the antenna port is $S_{33} = -24.8$~dB, indicating that the slot-line antenna is well matched to the guided mode, while the transmission from each waveguide port into the antenna feed is $S_{31} = S_{32} = -7.7$~dB. The low return loss confirms that the residual inefficiency arises at the guided-mode-to-feed transition rather than at the antenna, and the symmetric $S_{31} = S_{32}$ reflects the central detector placement.

The silicon waveguides were fabricated from commercial high-resistivity wafers ($\rho > 3~\mathrm{k\Omega\cdot cm}$, thickness $400$~$\mu$m) by photolithography and Bosch deep reactive-ion etching~\cite{Seliverstov2022}. The measured $S$-parameters of the bare structures confirmed insertion losses below $3$~dB across the operating band (Fig.~\ref{Fig1}(e)). The hBN-graphene-hBN heterostructure was then assembled by mechanical exfoliation and dry transfer~\cite{CastellanosGomez2014} directly onto the waveguide surface, with the graphene channel aligned to the antenna feed gap. Full fabrication details are provided in the supplementary material.

\begin{figure*}[ht]
  \centering
  \includegraphics[width=0.8\linewidth]{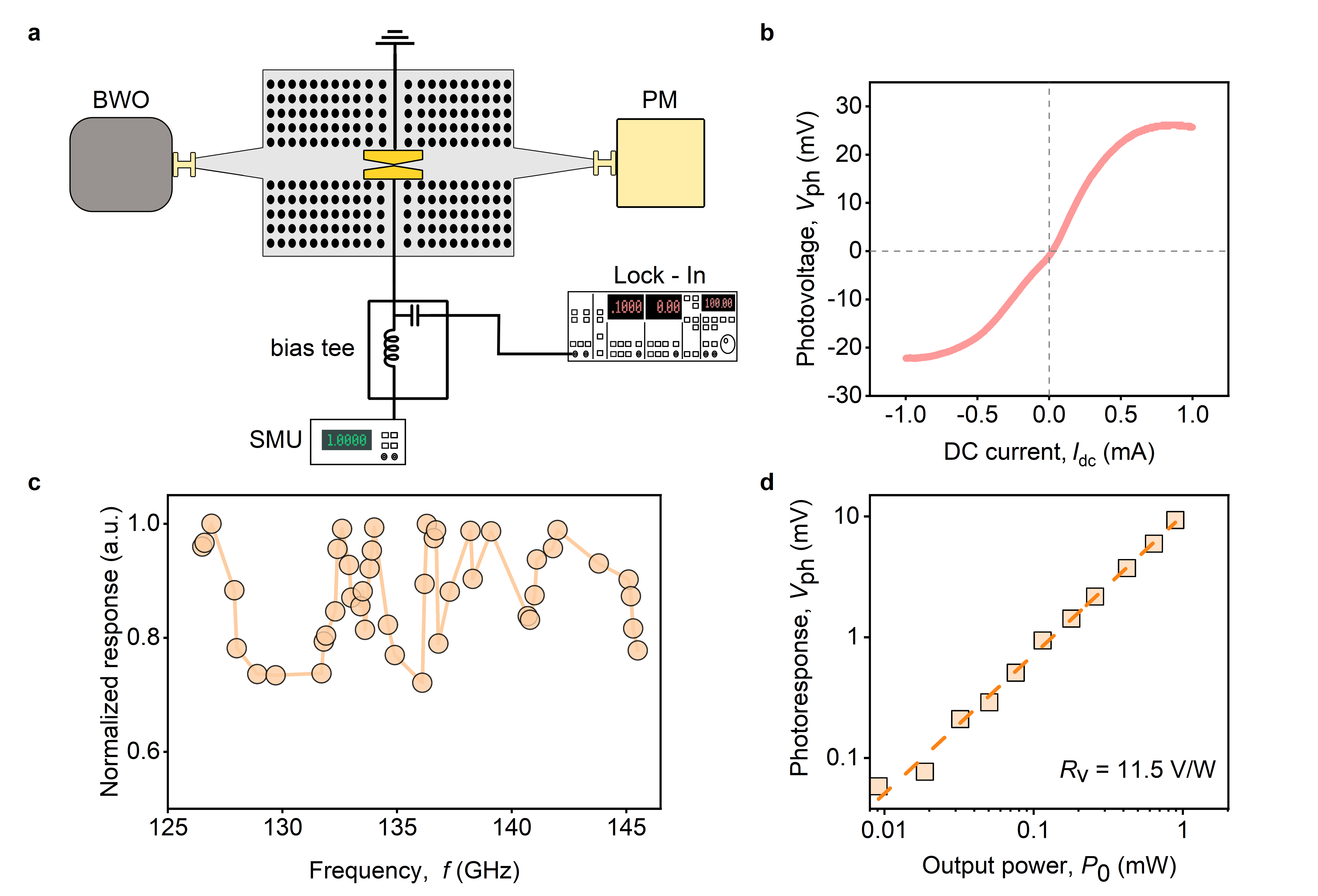}
  
  {\small
  \caption{
    \textbf{Terahertz photoresponse of the integrated graphene detector at room temperature.}
    \textbf{(a)} Schematic of the measurement setup. BWO output (amplitude-modulated 
    $100\,\mathrm{kHz}$) launched into silicon waveguide via WR-6 horn. Transmitted power 
    measured by calibrated PM. Graphene biased via SMU through bias tee and photovoltage 
    recorded by lock-in amplifier.
    \textbf{(b)} Photovoltage $V_{\mathrm{ph}}$ versus DC bias current $I_{\mathrm{dc}}$ at $f = 139\,\mathrm{GHz}$. Odd response in $I_{\mathrm{dc}}$ vanishing at zero bias is characteristic of bolometric detection. 
   \textbf{(c)} Normalized photovoltage across $125$--$146\,\mathrm{GHz}$ at fixed bias. The response varies by less than $30\%$ across the band.
    \textbf{(d)} Photovoltage versus incident power $P_\mathrm{0}$ (log-log). Linear scaling 
    $V_{\mathrm{ph}} \propto P_\mathrm{0}$ (slope $= 1$) confirms operation in the 
    small-signal regime. Voltage responsivity $R_\mathrm{V}$ = 11.5 V/W.
  }
  \label{Fig3}
  } 
\end{figure*}

We first investigated the transport characteristics of the fabricated detectors (Fig.~\ref{Fig2}). Two nominally identical devices were characterized. Data for the second detector are provided in the supplementary material. A representative current--voltage ($I$--$V$) curve at $T_\mathrm{0} = 300$~K is shown in Fig.~\ref{Fig2}(a). The curve is symmetric in bias polarity and weakly nonlinear. The deviation from ohmic behavior at higher $|I_{\mathrm{dc}}|$ is attributed to Joule self-heating of the channel. We next measured the temperature dependence of the differential resistance $R_\mathrm{d}$ over the range $313$--$333$~K (Fig.~\ref{Fig2}(c)). This dependence is well described by a linear function with a positive slope ($dR/dT > 0$), consistent with metallic-like transport at the operating Fermi level, which is set by the residual doping of the graphene. From this dependence, we extract the temperature coefficient of resistance as $\alpha = (1/R_\mathrm{0})(dR/dT) = 0.18\%~\mathrm{K}^{-1}$. This value is about one order of magnitude below the TCR of commercial vanadium-oxide or amorphous-silicon microbolometers, which reach $2$--$3\%~\mathrm{K}^{-1}$~\cite{ravindra2021microbolometers}. This gap is not fundamental. In high-quality hBN-encapsulated graphene the TCR is strongly enhanced near the charge-neutrality point, where the metallic-to-insulating crossover makes transport most temperature-sensitive, reaching values as high as $\sim 1\%~\mathrm{K}^{-1}$ at room temperature~\cite{yuan2020room}. Introducing an electrostatic gate to tune the Fermi level toward this regime is therefore expected to substantially increase $\alpha$.

We next extracted the thermal conductance $G$ of the graphene channel to its thermal sink from the self-heating contribution to the $I$--$V$ characteristic, which follows from the classical bolometric model~\cite{an2024high}:
\begin{equation}
    \frac{1}{R(I)} = \frac{1}{R_\mathrm{0}} - \frac{\alpha\,I^{2}}{G}.
    \label{eq:self_heat}
\end{equation}
Fitting Eq.~\eqref{eq:self_heat} to the experimental $\sigma(I_{\mathrm{dc}}) = 1/R_\mathrm{d}(I_{\mathrm{dc}})$ curve (Fig.~\ref{Fig2}(b)) with $\alpha$ fixed from $R_\mathrm{d}(T)$ yields $G = 3.5$~$\mu$W/K. Together, $\alpha$, $R_\mathrm{0}$, and $G$ set the expected bolometric voltage responsivity $R_{V} = I_{\mathrm{dc}}\,\alpha R_\mathrm{0}/G$, referenced to the power absorbed in the channel. At the operating bias $I_{\mathrm{dc}} = 800~\mu$A this evaluates to $\approx 250$~V/W. The responsivity measured in the following section, $R_{V} = 11.5$~V/W, is referenced instead to the power at the silicon waveguide port and is therefore lower by the power-coupling efficiency of the device. This efficiency is set mainly by the simulated antenna coupling $|S_{31}|^{2} = -7.7$~dB together with the incomplete absorption of the coupled power in the compact channel, and it accounts for the difference between the two values. The agreement in order of magnitude confirms that the measured photoresponse is consistent with the bolometric mechanism.

\begin{figure}[htbp]
    \centering
    \centering\includegraphics[width=0.85\linewidth]{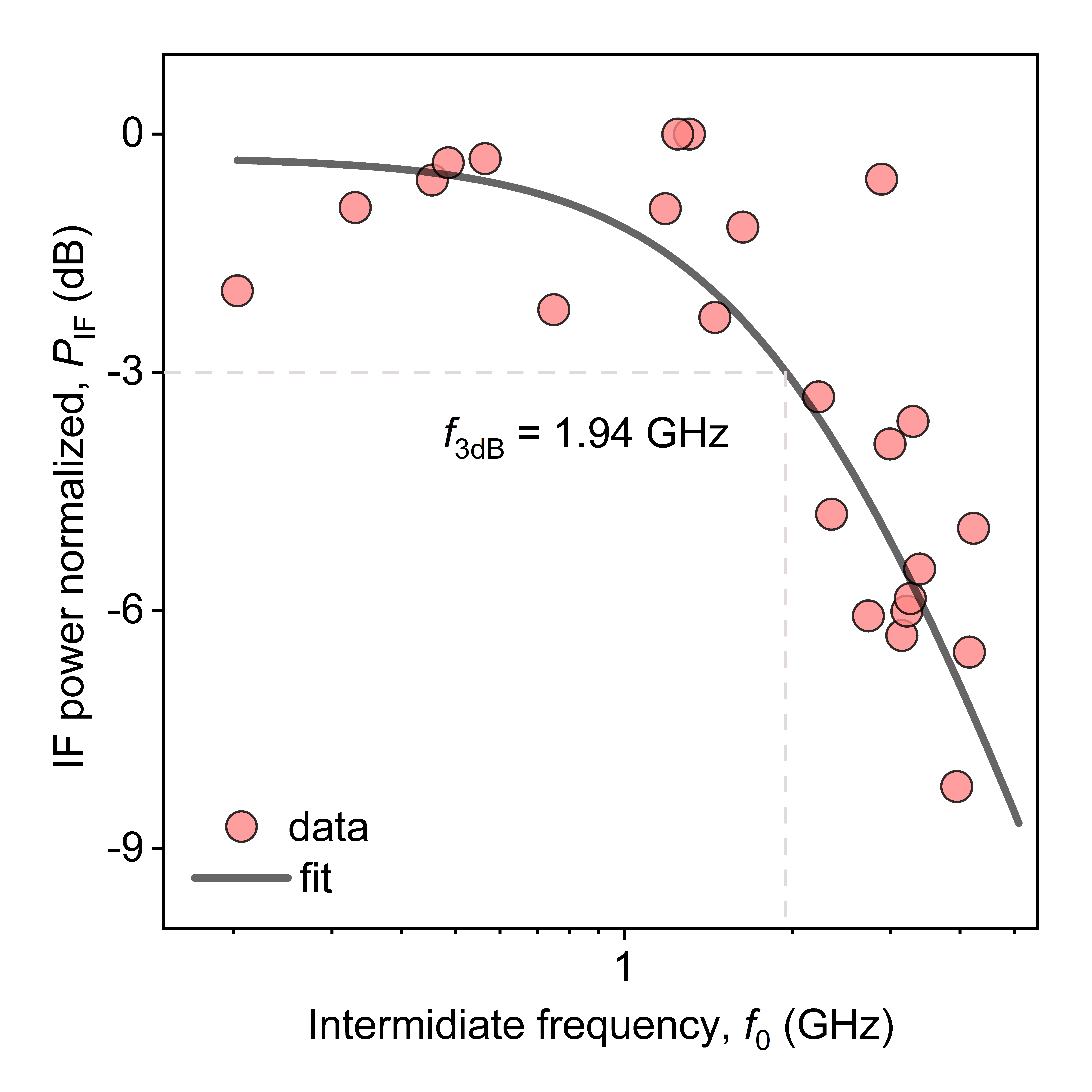}
   \caption{\textbf{Intermediate-frequency response of the graphene detector.}
    Normalized IF power $P_{\mathrm{IF}}$ as a function of the intermediate frequency $f_{\mathrm{IF}}$, measured by heterodyne mixing the BWO signal with a local oscillator and recording the down-converted signal at the graphene channel terminals. The solid line is a single-pole low-pass fit, yielding a 3 dB bandwidth $f_{3\mathrm{dB}} = 1.94$ GHz. This value agrees with the estimate of a parasitic bond-wire inductance.
    }
    \label{Fig4}
\end{figure}
The performance of the integrated graphene detector was characterized using the experimental setup shown in Fig.~\ref{Fig3}(a). Continuous-wave radiation from a backward-wave oscillator (BWO) was coupled into the silicon dielectric waveguide through a WR-6 horn and a tapered silicon transition. The BWO power and the waveguide insertion loss were calibrated using a power meter (see supplementary material). The graphene channel was biased through a bias tee connected to a source-measure unit, and the photovoltage was recorded by a lock-in amplifier synchronized to a $100$~kHz amplitude modulation of the BWO source. Figure~\ref{Fig3}(b) shows the photovoltage $V_{\mathrm{ph}}$ as a function of the DC bias current $I_{\mathrm{dc}}$ at $f_\mathrm{0} = 139$~GHz. The dependence is approximately odd in $I_{\mathrm{dc}}$, with $V_{\mathrm{ph}}$ vanishing near $I_{\mathrm{dc}} = 0$ and reversing sign with the bias polarity. At small bias the response is linear in $I_{\mathrm{dc}}$, while at higher $|I_{\mathrm{dc}}|$ it gradually saturates as Joule self-heating elevates the channel temperature and reduces the effective $\alpha\,\Delta T$ per unit current. This behavior is the canonical signature of bolometric detection in a current-biased configuration, in which $V_{\mathrm{ph}} = I_{\mathrm{dc}}\,\alpha R_\mathrm{0}\,\Delta T_{\mathrm{rad}}$ with $\Delta T_{\mathrm{rad}}$ the radiation-induced temperature rise of the graphene channel. The frequency dependence of the normalized photoresponse measured across $125$--$146$~GHz at fixed bias is shown in Fig.~\ref{Fig3}(c). The response varies by less than $30\%$ across the investigated band, confirming that the antenna-coupled detector operates without sharp resonances over the relevant portion of the D-band. We attribute the ripple to weakly localized modes at the transition between the tapered coupler and the slot-line antenna, associated with residual impedance mismatch in this region, rather than to an intrinsic feature of the detector. Figure~\ref{Fig3}(d) shows the dependence of $V_{\mathrm{ph}}$ on the BWO output power $P_{0}$ measured by the power meter. The photovoltage scales linearly with $P_{0}$ over more than two decades, confirming operation in the small-signal regime and ruling out higher-order rectification mechanisms. Combined with the odd bias dependence of Fig.~\ref{Fig3}(b), this linearity establishes the bolometric origin of the photoresponse. From the slope we extract a voltage responsivity $R_\mathrm{V}$ = 11.5~V/W, referenced to the power coupled into the silicon dielectric waveguide port. The noise-equivalent power, obtained from the output voltage noise measured with a Unipan~233 nanovoltmeter and referenced to the same waveguide port, is $\mathrm{NEP} = 2.46$~nW/Hz$^{1/2}$. These figures place the present waveguide-integrated detector within the range reported for free-space graphene sub-THz detectors~\cite{cai2014sensitive,Castilla2019,asgari2021chip}, while being obtained in a fully monolithic, waveguide-coupled geometry.

The detector response speed is a second key figure of merit for sub-THz communication receivers. Graphene-based detectors have been shown to support intrinsic bandwidths exceeding tens of GHz, with the measured performance typically limited by the external readout and packaging rather than by the graphene itself~\cite{titova2026fast,soundarapandian2026high}. The bandwidth of our device was characterized using a heterodyne measurement scheme employing two BWOs, one acting as the local oscillator (LO) and the other as the RF source. Details of the setup are given in the supplementary material. Figure~\ref{Fig4} shows the normalized intermediate-frequency power as a function of $f_{\mathrm{IF}} = f_{\mathrm{RF}} - f_{\mathrm{LO}}$. The response exhibits a single-pole roll-off with a $3$-dB cutoff at $f_{3\mathrm{dB}} = 1.94$~GHz. The graphene channel is connected to the signal extraction pads by ultrasonic bond wires and mounted on an FR4 carrier. The bandwidth is limited by the parasitic reactance of this readout, namely the inductance of the bond wires together with the capacitance of the contact pads and the FR4 substrate, which forms a low-pass network with the channel resistance $R_\mathrm{0} = 600~\Omega$. This limit is external to the detector, and the measured $f_{3\mathrm{dB}}$ therefore reflects the packaging rather than the intrinsic response of the graphene channel. Shorter bond wires, lower-permittivity carriers, and impedance-matched on-chip readout are expected to extend the bandwidth. 

We have demonstrated a room-temperature graphene sub-THz detector monolithically integrated on a high-resistivity silicon dielectric waveguide and operating over a 125-146~GHz span of the D-band. The active element, an hBN-encapsulated graphene channel coupled to the guided mode through a tapered slot-line antenna, exhibits a bolometric photoresponse. The device delivers a voltage responsivity $R_\mathrm{V}$ = 11.5~V/W referenced to the power coupled into the silicon waveguide port, and a $3$-dB bandwidth $f_{3\mathrm{dB}} = 1.94$~GHz. The measured bandwidth is fully accounted for by the $L/R$ cutoff of the wire-bonded readout and is therefore set by external parasitics rather than by the intrinsic response time of the graphene channel~\cite{titova2026fast}. Two routes are identified for further improvement. First, introducing an electrostatic gate would allow the Fermi level to be tuned toward the near charge neutrality point, where $|dR/dT|$ is maximized, enabling a substantial increase of $R_\mathrm{V}$. Second, replacing wire bonding with impedance-matched on-chip readout is expected to push the bandwidth well into the tens-of-GHz range relevant to $6$G receivers. Together with the monolithic, room-temperature character of the platform, these results establish a scalable route toward fully integrated sub-THz photonic circuits combining low-loss dielectric waveguides with active graphene elements.
\section*{Supplementary Material}
See the supplementary material for the silicon waveguide 
fabrication route and hBN-encapsulated detector processing 
details, the scalar power-based $S$-parameter measurement 
procedure, and the electrical and photoresponse 
characterization of a second waveguide-integrated detector.

\begin{acknowledgments}
This work was supported by the Russian Science Foundation (Grant No.\ 23-72-00014) and the Basic Research Program of HSE University (HSE-BR-2025-002).
\end{acknowledgments}

\section*{Author Declarations}
\subsection*{Conflict of Interest}
The authors have no conflicts to disclose.

\subsection*{Author Contributions}
\textbf{A. Titchenko:} Conceptualization (equal); Investigation (equal); 
Writing -- original draft (lead).
\textbf{K. Shein:} Investigation (supporting); Writing -- review \& editing (equal).
\textbf{M. Titova:} Investigation (supporting); Resources (supporting).
\textbf{M. Kashchenko:} Investigation (supporting).
\textbf{O. Popova:} Investigation (supporting).
\textbf{R. Izmaylov:} Investigation (supporting).
\textbf{E.I. Titova:} Resources (supporting); Supervision (supporting).
\textbf{I. Gayduchenko:} Conceptualization (equal); Supervision (lead); 
Writing -- review \& editing (equal).
\textbf{G. Goltsman:} Funding acquisition (lead); Supervision (supporting).

\section*{Data Availability Statement}
The data that support the findings of this study are available 
from the corresponding author upon reasonable request.

\bibliographystyle{aipnum4-1}     
\bibliography{references}
\end{document}